\documentclass[aps,prl,floats,showpacs,twocolumn]{revtex4-1} 
\usepackage{color}
\usepackage{amsmath}
\usepackage{hyperref}
\usepackage{graphicx}

\def\be{\begin{equation}}
\def\ee{\end{equation}}
\def\bea{\begin{eqnarray}}
\def\eea{\end{eqnarray}}

\def\sss{\scriptscriptstyle}
\def\vk{\mathbf{k}}
\def\EF{E_{\sss F}}

\def\mf{m_{\sss M}}
\def\nf{n_{\sss M}}
\def\muf{\mu_{\sss M}}
\def\gf{g_{\sss M}}
\def\wg{\sss WG}
\def\etil{e_{\sss X}}
\def\alptil{\alpha_{\sss X}}

\begin{document}
\title{Dark Energy from ferromagnetic condensation of cosmic magninos}
\author{Urjit A. Yajnik}
\email{yajnik@iitb.ac.in}
\affiliation{Department of Physics, Indian Institute of Technology, 
Bombay, Mumbai 400076 India}
\affiliation{Department of Physics, McGill University, Montr\'eal QC H3A 2T8, Canada} 
\date{26 January 2011}

\begin{abstract}
It is proposed that an ultra-light fermionic species, dubbed cosmic magnino 
has condensed into a  ferromagnetic state in the Universe. The extended structure 
of domain walls associated with this ferromagnetism accounts for the observed 
Dark Energy. In modification of the situation with an electron gas, it is proposed that 
the Stoner criterion is satisfied due to magnetic dipolar repulsion. 
The cosmological requirements then yeild a lower bound on the magnetic moment 
of the cosmic magnino. The proposed magnetism is supposed to be associated with a new 
non-standard  electromagnetism. If the magnino is also electrically charged under 
this electromagnetism,  the corresponding oppositely charged heavier species would 
account partially or entirely for the Dark Matter in the Universe. 
\end{abstract}

\maketitle
\section{Introduction}
The discovery of the Dark Energy component \cite{Riess:1998cb}\cite{Perlmutter:1998np}
of the cosmological energy density from direct observations, in concordance with the
WMAP precision data \cite{Komatsu:2010fb} presents a new challenge to fundamental 
physics. The extremely small value of the mass scale associated with this energy density
makes it unnatural as Cosmological Constant\cite{Weinberg:1988cp}, and therefore
demands an unusual mechanism for relating it to the known physics of
elementary particles. On the other hand a new window to the very low mass physics 
has been opened up by the discovery of the low mass scale of neutrinos \footnote{For reviews,
se \cite{Bahcall:2004ut}, \cite{Maltoni:2004ei}}.
Further, a variety of theoretically motivated ultra-light species are currently
being sought experimentally \cite{Jaeckel:2010ni}. We may therefore exploit the
presence of an ultra-light sector to explain the Dark Energy phenomenon autonomously 
at a low scale, without direct reference to its high scale connection with known physics.

The analysis of \cite{Komatsu:2010fb} assumes the $\Lambda$-CDM model, with  equation 
of state of the Dark Energy constrained to $p/\rho\equiv w=-1$. However alternative 
analyses (see for instance   \cite{Shafieloo:2009ti}) show that dynamically evolving $w$ 
is alsoconsistent with data. 
It has been argued, early in \cite{Battye:1999eq} that the equation of state 
obeyed by the observed contribution to the energy density could be well fitted by a 
network of frustrated domain walls \cite{Kibble:1980mv}, which obey an effective 
equation of state $p=(-2/3)\rho$ in the static limit.  
This  possibility has been further examined 
in \cite{Conversi:2004pi}\cite{Friedland:2002qs} and shown to be still consistent 
with more recent data. 
Here I adopt the approach that the most robust feature of the astrophysical 
obsevations, namely $p/\rho<0$ at redshift $z\approx 0.3$ can be fitted by space 
filling extended objects. The extended structure is generated by a specific mechanism 
involving strongly correlated  but weakly  coupled dynamics, which leads to definite 
predictions about masses and cosmic history of some of the hidden contents of the Universe. 

The proposed model involves a new fermionic species responding to a hidden electromagnetism 
and whose condensed state can be characterised 
by the familiar Stoner criterion \cite{Stoner} of ferromagnetism. A novel feature 
I propose \cite{Yajnik:2005pr} is to assume the Stoner 
mechanism to arise from dipolar force between magnetic moments. 
The reason is that the fermion gas in the cosmological setting has to be extremely 
rarefied in which case the screened  Coulomb potential which is exponentially cut 
off becomes ineffective, while magnetic force is not screened and obeys a power law. 
The proposed mechanism of condensation due to magnetic dipolar interaction between 
neutral fermions was first propsed in \cite{Yajnik:2005pr}, which led to the 
calculations of \cite{0305-4470-39-6-002}\cite{PhysRevE.76.062101} 
\footnote{This has subsequently received attention in \cite{PhysRevLett.103.205301}}. 
Stoner ansatz has received extensive theroetical attention in the past decade. 
In the context of a three dimensional gas, controlled verification of Stoner mechanism 
for neutral ultra-cold gas of spinless atoms is reported in \cite{Jo18092009}, where 
a Feshbach  resonance is used for tuning the repulsion. 
It would be interesting to also test the dipolar mechanism proposed here in 
a laboratory system.

We dub this species \textit{magnino}, being an ultra-light fermionic species whose predominant
property manifesting itself today is magnetism. As to the magnetic moment, there are 
two possibilities.
One is that the magnetic moment is induced, and the other that it is intrinsic.
In the latter case, the magnino is also electrically charged, and another species, 
equally and oppositely charged should be present for neutrality of the Universe. 
This gives rise to the further possibility that it  is heavier, not  participating 
in the condensation mechanism, but be a component of  Dark Matter.

The proposed new species, plus one or more accompanying species such as the corresponding
photons, would contribute to the tally of relativistic species 
present at the epoch of Big Bang Nucleosynthesis ( BBN). This is justified by the analyses
of new data \cite{Komatsu:2010fb} and \cite{Dunkley:2010ge} which admit the presence of
approximately one new effective degree of freedom.
The proposed model can accommodate a  Dark Matter candidate, in which case
it would also provide an explanation for the ``cosmic concordance''. 
An additional possibility in the mechanism is an explanation of the seeds required 
to produce the  inter-galactic magnetic fields, as I discuss at the end. 
In the following I use the 
units $\hbar=c=1$ such that $\hbar c=1\approx 200$MeV-fermi, and express all dimensionful  
quantities in the units of eV.

\textit{Cosmological setting :} 
A gas of slowly moving thin domain walls has an average stress tensor 
given by \cite{KolTur} 
\be
\label{eq:Twalls}
\langle T^\mu_\nu \rangle = \frac{\eta^3}{3L} \textrm{diag}\{3,-2,-2,-2 \}
\ee
where $\eta$ is the mass scale associated with the surface tension of the walls,
and $L$ is the average separation between the walls. Thus the wall gas (WG) satisfies 
the equation of state $p_{\wg}=(-2/3)\rho_{\wg}$.
A Universe containing non-relativistic matter,  and domain wall 
structure formed on scales much smaller than the Hubble scale, is described by 
Friedmann-Robertson-Walker scale   factor $R(t)$ obeying
\be
\label{eq:scalefactor}
\left( \frac{1}{R}\frac{dR}{dt} \right)^2 = \frac{8\pi}{3} G 
\left( \frac{R_0^3 \rho_{m0}}{R^3} + \frac{\rho_{\wg0}R_0}{R} \right)
\ee
where the subscript $0$ refers to the present epoch.  Let $t_1$ be the time when the 
the energy density of the walls gas equals the energy density in non-relativistic matter. 
Using the value of density fraction of matter $\Omega_m\approx 0.3$ and that of 
Dark Energy $\Omega_\Lambda\approx 0.7$ at the present epoch, gives $(R_1/R_0)^2=3/7$.  
Photon temperature at this epoch is  $T_1=4.18 K= 5.0\times 10^{-4}eV$. The current 
contribution of Dark energy to the total energy density has the 
value $\rho_{\scriptstyle{DE}}\approx (3\times10^{-3}\textrm{eV})^4$.

We shall assume the magnino (M) to be a very light species, of mass $\mf \leq T_1$,
with a conserved number so that the abundance of the species 
relative to photons remains constant during the epochs under consideration.
The number density of the species
can be parameterised as $\nf(t_1)\ =\ 3.56\times 120 \Upsilon {\rm cm}^{-3}
\approx 3.2 \times 10^{-12} \Upsilon (eV)^3$, 
where $\Upsilon$ is an unknown factor. 

\textit{Ferromagnetism:} According to the Stoner ansatz \cite{Stoner} 
spontaneous ferromagnetism is a consequence of a shift in single particle 
energies, proportional to the  difference between
the spin up ($N_{\uparrow}$) and the spin down ($N_{\downarrow}$) populations. 
A parameter $I$
is introduced to incorporate this, the single-particle energy spectrum being
\be
\label{eq:Esplit}
E_{\uparrow, \, \downarrow}(\vk) = E(\vk) - I 
\frac{N_{\uparrow, \, \downarrow}}{N}
\ee
Using this it is shown \cite{Brout}\cite{IbLu} 
that the ferromagnetic  susceptibility is
\be
\label{eq:suscep}
\chi\ =\ \frac{\chi_{\sss P}}{1 - I \frac{\beta}{\EF}}\ 
\ee
where $\beta$ is a factor of order unity depending upon the geometry of
the Fermi surface; for the spherical case having value $\frac{3}{4}$. 
The criterion for spontaneous magnetization is $\chi<0$. A sufficient
condition  for the gas to be spontaneously magnetised at zero temperature  
is the Stoner criterion,
\be
\label{eq:stoner}
I\ >\ \frac{\EF}{\beta}
\ee

The origin of such a large energy shift is supposed to be a repulsive
interaction among the fermions, which makes it favourable for them to
enter the state of aligned spins, which in turn due to Pauli exclusion principle
ensures large enough a separation among them so as to reduce the 
repulsive energy. 
An estimate of the size of this ``exchange hole''
 \cite{FetWal}\cite{IbLu} is given by the density deficit of same spin fermions
in the vicinity of a given fermion, $\Delta \nf\  =\   -0.86\nf$. Let the 
two-particle long range interaction  energy be $\gamma^2$ which is repulsive.   
This energy reduction should be proportional to $\Delta \nf$. For the Stoner 
parameter I therefore stipulate the relation
\be
I = \gamma^2 \frac{\vert \Delta \nf\vert}{\nf }
\label{eq:Idef}
\ee
I now make the assumption that for the fermions under consideration, this 
coupling arises from magnetic dipole-dipole interaction, which is dominated
by a repulsive contribution in an appropriate ferromagneic state. The resulting 
increase in single particle energy can be estimated  as 
\be
\label{eq:coupling}
\gamma^2\ =\  \kappa_{\scriptscriptstyle{JM}} \muf^2 |\Delta \nf|
\ee 
The favorable ferromagnetic state, the JM ansatz \cite{PhysRevE.76.062101} turns 
out to be spheroidal, ensuring dominance of repulsion, and the related 
parameter $\kappa_{\scriptscriptstyle{JM}}$ is a factor of order unity.
Note that the interaction energy between non-relativistic dipoles goes as inverse 
third power of interparticle separation and hence consistent with scaling as $ |\Delta \nf|$.

The magnetic moment introduced above could either be intrinsic or induced, 
\be
\label{eq:mudef}
\muf \equiv \gf\frac{\etil\,\hbar}{2\mf}  
\ee
where $\etil$ is the unit of charge of the new electromagnetism, and $\mf$
is the mass of the magnino. For charged magnino, $\gf$ at tree level has the Dirac value $2$. 
If the magnino is to be electrically neutral, the factor $\gf$ has to arise 
from radiative corrections or from compositeness.
For example, for neutrinos the radiatively induced magnetic moment is expected 
to be small \cite{Marciano:1977wx}, or $\muf/\mu_{\sss B}<10^{-15}$ 
as derived in \cite{Bell:2005kz} under certain reasonable assumptions. 
In a more general setting, $\gf$ can be order unity as in the case of the neutron. 
Thus the Stoner criterion (\ref{eq:stoner}) becomes
\be
\label{eq:stonercosmic}
\alptil \nf \left(\frac{\gf}{\mf}\right)^2
> \frac{4}{3} \left\{\left((3\pi^2\nf)^{2/3} +\mf^2\right)^{1/2} -\mf\right\}
\ee
where $\alptil=\etil^2$ is the fine structure constant, and we have assumed $|\Delta \nf|\approx \nf$, 
and $\kappa=1$ for simplicity. In the non-relativistic approximation when this mechanism
is assumed to operate, this requirement can be simplified to $\mf < \frac{1}{6}\alptil \nf^{1/3} $. 
The critical temperature for such a phase transition is estimated to be  $T_c=I /4$\cite{IbLu}. 
In the present case, this value becomes $T_c\simeq \mf/{\alptil}^2$.
This is far too high, since it allows creation of magnino pairs and condensation cannot be stable.
We assume in the following that there is an epoch $t_2$ preceding $t_1$ when this phase transition
is accomplished, with the requirement that the temperature $T_{h2}$ of the hidden photons
at $t_2$ satisfies $T_{h2}<\mf$.

\textit{Domain walls:} 
The energy density of the condensed state can be separated into a sum of two contributions,
one from the configuration  space degrees of freedom and the other due the spin degrees 
of freedom.  The 
latter may be modelled by
a Landau-Ginzberg lagrangian for the magnetisation $\bf M$, a vector  order parameter,
with a symmetry breaking self-interaction $\lambda 
({\bf M}\cdot{\bf M}-\sigma^2)^2$. Here $\sigma$ determines the magnitude 
of the magnetization per unit volume, estimated to be $\muf|\Delta \nf|$ \cite{IbLu} upto a 
factor of order unity. From standard solitonic calculation 
\cite{Rajaraman:1982is} the domain walls have a width $w\sim(\sqrt{\lambda}\sigma)^{-1}$ 
and energy  per unit area $\eta^3 \sim \sqrt{\lambda}\sigma^3$. These domain walls are
not expected to be topologically stable. This is because the vacuum manifold which is
a 2 dimenional sphere allows for the wall to develop holes, though classically suppressed 
by an energy barrier. 
The rate of decay of the walls is thus governed by tunneling processes \cite{Preskill:1992ck} 
which can be much slower than the age of the Universe.  
According to the JM ansatz for the ferromagnetic state introduced in \cite{PhysRevE.76.062101} 
the fermi surface of the condensate is spheroidal, leading also to ferronematic order 
\cite{PhysRevLett.103.205301} in configuration space. 

As for the momentum space degrees of freedom,
they continue to behave like a degenerate quantum gas; however the formation
of the domains breaks their almost scale invariant behaviour. Let us assume the
average size of the domains to be characterised by a length scale $l$.
Equivalently, there is one wall passing through a cubical volume 
of size $l^3$ on the average. The energy density trapped in such a wall is $\eta^3/w$. 
The scale $l$ of this microstructure is expected to be many orders of magnitude smaller
than the scale of galactic clusters, $l_{gc}$, and certainly the Hubble scale.
In the following we assume that the momentum degrees of freedom which are
$O(l_{gc}^{-1})$ and larger, and certainly those that are $l^{-1}$ and larger
cease to respond to the cosmic expansion. Only the  degrees of freedom of very low 
momentum, smaller than $l^{-1}_{gc}$, continue to redshift as $R^{-1}$.
The corresponding separation of the number densities may be denoted ${\nf}_>$
for the large momentum, which would be most of the number density contribution, 
and a small fraction ${\nf}_<$ for the low momentum modes.

This separation of scales makes it clear how the microscopic relation $\sigma \propto \nf$,
could be consistent with the fact that the coarse grained wall gas energy density obeys
the  equation of state $p_{\wg}=(-2/3)\rho_{\wg}$. Once the walls have formed, $\sigma$ 
is essentially
determined by ${\nf}_>$ and remains a constant. I assume therefore that the density 
$\rho_{\wg}$ is set by the initial value $\sigma^2(t_2)$ at the epoch $t_2$ of 
emergence of wall gas, and subsequently scales as $1/R$ as required by the covariant 
conservation equation $d(\rho_{\wg}R^3)+p_{\wg}dR^3=0$.  With this caveat, 
if the wall gas is to comprise the entire Dark Energy component of the Universe,
we get the relation
\be
\label{eq:mfcosmobound}
\rho_{\wg}\approx
\left( \frac{{\gf}^2\alptil}{\mf^2} \right) ({{\nf}_>(t_2)})^2 \left\langle 
\frac{w}{l}\right\rangle \frac{R_2}{R_0} 
\approx \rho_{\scriptstyle{DE}}
\ee
where we have replaced $w/l$ by its average value. 
The ratio $R_2/R_1\approx R_2/R_0$ requires the details of the ferromagnetic phase 
transition. If we assume the emergence of wall gas to have been not much earlier
than when the equality $\rho_{WG}\approx \rho_m$ was reached, then using $\gf^2\sim O(1)$, 
$\alptil\sim 10^{-2}$, and $\langle w/l \rangle \lesssim 10 $, we get a
bound $\mf/\Upsilon \lesssim 10^{-8}$eV. 

\textit{Other dark components:} If $\gf$ arises from generic radiative 
corrections, it is proportional to $\mf$ 
and then the above  scenario does not work. The possibility of 
the magnino being a neutral composite like the neutron remains open. A more appealing 
possibility is that it is charged, however the large intrinsic magnetic
moment could not be of standard electromagnetism without being
detected so far in some of the astrophysical phenomena such as the 
cooling rates of supernovae and red giants \cite{Raffelt:1990yz}. Thus the corresponding 
electromagnetism must be new. Then an oppositely charged partner 
must also be present in the Universe for neutrality. The results derived so
far remain unaffected if this partner is a heavier species, not participating 
in the ferromagnetism, in other words, if the new unobserved sector is also
asymmtric under charge conjugation like the observed one. Let us designate  this 
oppositely charged partner $Y$, and assume it to have equal and opposite charge, and
therefore the same abundance as, the magnino. 

The third new component, the photons of the new electromagnetic force should 
also be present, with an entropy density in a ratio $\Upsilon_\gamma$ 
to the entropy density of the standard photons.  
Since $T_{h2}<\mf$ at an epoch when usual photon have a temperature $T\gtrsim 10^{-4}$eV,
$\Upsilon_\gamma\ll 1$. 
From the observational bounds on the new effective relativistic degrees of freedom, 
it is necessary that $2\Upsilon + \Upsilon_{\gamma}<1$.
If we assume $Y$ to be a massive non-relativistic species at present epoch, we can now 
obtain a bound on its mass.
Using the standard values \cite{PDG} of relative density fractions
of baryonic and non-baryonic matter and the baryon to photon ratio, we get
$m_Y < (3.22/\Upsilon)$ eV in order for $Y$ to not overclose the Universe. Since 
$\Upsilon<\frac{1}{2}$,
$Y$ turns out to be non-relativistic at cosmic tempratures $<1$eV, i.e., after most of
the primordial neutral Hydrogen has come into existence. The mass range suggested
seems too light to act as Dark Matter to assist structure formation, however 
there is no model independent 
lower bound on the Dark Matter candidate \cite{Boyarsky:2008ju, Boyarsky:2008xj}, 
and the possibility of Dark Matter from hidden world can be analysed along the lines 
of \cite{Ignatiev:2003js}. Note that depending on the temperature $T_2$ to be deduced from
details of the wall codensation mechanism, $\Upsilon$ can be considerably smaller than unity, 
and $m_Y$ can then be sufficiently large for it to be acceptable Cold Dark Matter. 

Another interesting outcome of the magnetic nature of this condensation mechanism 
is that it may explain the existing inter-galactic  magnetic fields. For this it 
is necessary that the two electromagnetisms mix through kinetic terms. Then over 
a large number of domains 
encompassing the scale of galactic clusters, the fluctuations from the
average value zero of the net magnetic field may be large enough, that 
after mixing with standard electromagnetism it gives rise to the seeds
required for generating the intergalactic magnetic fields \cite{Kulsrud:1999bg}\cite{Kulsrud:2007an}.
These effects are the subject of ongoing work.

A characteristic prediction of  this mechanism is degradation of Dark Energy. 
The form of the Stoner requirement Eq. (\ref{eq:stonercosmic}) 
means that as the Universe expands, the left hand side of the inequality diminishes faster 
than the right  hand side, and eventually
the inequality cannot be satisfied. Thus sometime before tempreature $T\rightarrow0$,
the ferromagnetic state melts away, so does the wall gas masquarading 
as Dark energy, and the Universe returns to being matter dominated.

In conclusion I have introduced the cosmic magnino whose mass scale is
extremely small, perhaps to be matched only by that of the lightest neutrino.
Such light Dirac mass \cite{Witten:1979nr} as also the extra $U(1)$ gauge symmetry 
can be obtained in  the context of grand unification. 
Compositenss is also a possibility for the magnino, however ensuring a very
small mass and a large induced dipole moment may be difficult to arrange 
in this case. The easier alternative is for it to also be charged under an 
abelian gauge force. In this case the particle is further accompanied by a heavier
oppositely charged partner $Y$, which can potentially be Cold Dark Matter.

I have benefited from critical remarks by Guy Moore, H. B. Nielsen, Wang-Yi, 
Sandip Trivedi, Surjeet Rajendran and Bret Underwood and from extensive discussion 
with and encouragement  from S. S. Jha. 
It is a pleasure to thank the McGill High Energy Physics group for hospitality,
and for financial support during a sabbatical leave. 
This work is also supported by a grant from Department  of Science 
and Technology, India. 

%

\end{document}